\documentclass[prl,twocolumn,showpacs]{revtex4}
\usepackage{graphicx}
\usepackage{amsmath}
\usepackage{amssymb}

\def\lsim{\
  \lower-1.2pt\vbox{\hbox{\rlap{$<$}\lower5pt\vbox{\hbox{$\sim$}}}}\ }
\def\gsim{\
  \lower-1.2pt\vbox{\hbox{\rlap{$>$}\lower5pt\vbox{\hbox{$\sim$}}}}\ }

\begin{document}
\title{On the nature of the lowest state of a Bose crystal}
\author{Maksim D. Tomchenko} \email{mtomchenko@bitp.kiev.ua}
\affiliation{Bogolyubov Institute for Theoretical Physics, 14-b,
Metrolohichna Street, Kyiv 03143, Ukraine}

\date{\today}
\begin{abstract}
As is known, the ground state (GS) of a system of spinless bosons
must be non-degenerate and must be described by a nodeless wave
function. With the help of the general quantum mechanical analysis
we show that any \textit{anisotropic} state of a system of spinless
bosons is degenerate. We prove this for a two-dimensional (2D)
system, infinite or finite circular, and for a three-dimensional
(3D) system, infinite or finite ball-shaped. It is natural to expect
that this is valid for finite 2D and 3D systems of any shape. Hence,
GS of a Bose system of any density is isotropic and, therefore,
corresponds to a liquid or  gas. Therefore, the lowest state of a 2D
or 3D natural crystal consisting of spinless bosons should be
described by a wave function with nodes. This leads to nontrivial
experimental predictions. We propose a possible ansatz for the wave
function of the lowest state of a 3D Bose crystal and discuss
possible experimental consequences.
\end{abstract}

%\textbf{Keywords:} Bose crystal;  ground state;  degeneracy. \\ \\

 \maketitle

Experiments show that the cooling of a Bose liquid and the
compression of a Bose gas of low temperature cause the
crystallization. This is seen from the $(P,T)$ phase diagrams of
inert elements \cite{glazov,leachman,pavese}. The only exception is
helium. This enables one to make the obvious (at first glance)
conclusion that, at $T=0,$ a crystal is the genuine (nodeless)
ground state (GS) of a Bose system
\cite{guyer,leggett2006,cazorla2017}. In this Letter we will show
that the genuine GS of a two-dimensional (2D) and three-dimensional
(3D) infinite system of spinless bosons of any density is isotropic
and, therefore, corresponds to a liquid or gas. Moreover, we will
briefly discuss the meaning of this from the experimental point of
view.

The proof is simple. Consider the infinite 3D system ($N,
V\rightarrow\infty$, $N/V=const$) of $N$ interacting spinless
bosons. The uniformity  and the isotropy of the space ensure the
invariance of the Hamiltonian of the system
\begin{equation}
 \hat{H}_{\textbf{r}} = -\frac{\hbar^{2}}{2m}\sum\limits_{j=1}^{N}\triangle_{\textbf{r}_{j}} + \frac{1}{2}\sum\limits_{j,l}^{l \not= j}
 U(|\textbf{r}_{l}-\textbf{r}_{j}|)
     \label{1-1} \end{equation}
with respect to translations
$\textbf{r}_{j}\rightarrow\textbf{r}_{j}+\textbf{a}$ and rotations
$\textbf{r}_{j}\rightarrow\acute{\textbf{r}}_{j}=A\textbf{r}_{j}$
\cite{land3,vak} (here, $j=1,\ldots,N$,  $A$ is the rotation matrix,
and the vectors $\textbf{r}_{j}$ and $\acute{\textbf{r}}_{j}$ are
given in the same basis). Hence, $\hat{H}$ commutes with the
operators of translations
$\hat{\textbf{T}}=e^{i\textbf{a}\textbf{P}/\hbar}$  and rotations
$\hat{R}=e^{i\varphi \textbf{i}_{\varphi}\textbf{L}/\hbar}$
\cite{land3,vak,petrashen}. Therefore, $\hat{H}$ commutes also with
the operator of total momentum $\hat{\textbf{P}}=-i\hbar
\sum_{j=1}^{N}\frac{\partial}{\partial \textbf{r}_{j}}$ and the
operator of total angular momentum
$\hat{\textbf{L}}=-i\hbar\sum_{j=1}^{N}
[\textbf{r}_{j}\times\frac{\partial}{\partial\textbf{r}_{j}} ]$. As
is known from quantum mechanics, in this case there must exist a
common set of eigenfunctions of the operators $\hat{H}$,
$\hat{\textbf{L}}^{2}$ and $\hat{L}_{z}$, and a common set of
eigenfunctions of the operators $\hat{H}$ and $\hat{\textbf{P}}$
\cite{land3,vak} [the boundary conditions (BCs) have to admit such
properties; therefore, we consider the \textit{infinite} system;
since the space is uniform and isotropic, BCs at infinity are
invariant under translations and rotations: this is obvious for a
nonclosed system and can be shown for a closed one]. These sets of
functions are different, because $\hat{\textbf{L}}$ and
$\hat{\textbf{P}}$ do not commute:
$[\hat{L}_{\alpha},\hat{P}_{\beta}]=i\hbar \sum_{\gamma} e_{\alpha
\beta \gamma}\hat{P}_{\gamma}$, where $\alpha, \beta, \gamma=x, y,
z$, $e_{xyz}=e_{zxy}=e_{yzx}=1$, $e_{yxz}=e_{zyx}=e_{xzy}=-1$, and
$e_{\alpha \beta \gamma}=0$ for the rest ones $\alpha, \beta,
\gamma$ \cite{land3}.

Consider the common set of eigenfunctions of the operators $\hat{H}$
and $\hat{\textbf{P}}$. Each state with nonzero momentum is
degenerate. Indeed, since
$[\hat{\textbf{L}}^{2},\hat{P}_{\beta}]\neq 0$, the function
$\hat{\textbf{L}}^{2}\Psi$ does not coincide with $const\cdot \Psi$
for any eigenfunction $\Psi$ of the operator $\hat{\textbf{P}}$. On
the other hand,
$\hat{H}\hat{\textbf{L}}^{2}\Psi=\hat{\textbf{L}}^{2}\hat{H}\Psi=E\hat{\textbf{L}}^{2}\Psi$.
That is, the level $E$ is at least twice degenerate, since two
different wave functions (WFs), $\Psi$ and
$\hat{\textbf{L}}^{2}\Psi$, correspond to the energy $E$
\cite{land3}. The states $\Psi_{j}$ with zero momentum
($\hat{\textbf{P}}\Psi_{j}= \textbf{P}\Psi_{j} =0$) should be
considered separately, since for them the commutation relations
$[\hat{\textbf{L}}^{2},\hat{P}_{\alpha}]=i\hbar
(\hat{P}_{\beta}\hat{L}_{\gamma}+\hat{L}_{\gamma}\hat{P}_{\beta})
-i\hbar(\hat{P}_{\gamma}\hat{L}_{\beta}+\hat{L}_{\beta}\hat{P}_{\gamma})$
 [where $(\alpha,\beta,\gamma)$ is $(x,y,z)$, or $(y,z,x)$, or
$(z,x,y)$] admit $\hat{\textbf{L}}^{2}\Psi_{j} =const\cdot
\Psi_{j}$. It is clear from the physical reasoning that GS WF of a
crystal, $\Psi_{0}^{c}$, should correspond namely to the zero
momentum. Since $\hat{\textbf{P}}^{2}$ commutes with $\hat{H}$,
$\hat{\textbf{L}}^{2}$ and $\hat{L}_{z}$ \cite{kirz}, the complete
set of WFs can be constructed so that they are eigenfunctions of the
operators $\hat{H}$, $\hat{\textbf{L}}^{2},$ $\hat{L}_{z}$ and
$\hat{\textbf{P}}^{2}$. In this case, on the states with
$\textbf{P}^{2}= 0$  the operator $\hat{\textbf{L}}^{2}$ must give
the definite values: $\hat{\textbf{L}}^{2}\Psi=\hbar^{2} L(L+1)\Psi$
with $L= 0$ or $L\geq 1$. The crystal is anisotropic, and the
anisotropy of a state means that there exists the rotation for which
$\hat{R}\Psi(\textbf{r}_{1},\ldots,\textbf{r}_{N})\neq
\Psi(\textbf{r}_{1},\ldots,\textbf{r}_{N})$. In this case,
$\hat{\textbf{L}}\Psi\neq 0$, because $\hat{R}=e^{i\varphi
\textbf{i}_{\varphi}\textbf{L}/\hbar}$. Hence, for the crystal
states with $\textbf{P}^{2}= 0$ we must have $L\geq 1$. Then the
state $\Psi_{0}^{c}$ is  $(2L+1)$-fold degenerate due to the
noncommutativity of $\hat{L}_{x}$, $\hat{L}_{y}$, and $\hat{L}_{z}$
\cite{land3,vak}.

We have shown for an infinite 3D system that any state with nonzero
momentum is degenerate, and among the states with zero momentum only
the isotropic state ($L= 0$) is non-degenerate.

Note that \textit{any} anisotropic state of an infinite 3D system
with energy $E$ is degenerate. Indeed, WF of such a state can be
expanded in WFs with definite squared angular momentum
$\hat{\textbf{L}}^{2}$ and with the same energy $E$. Such an
expansion should contain at least one WF  corresponding to $L\geq
1$, which leads to the degeneracy.

% In this case, WF $\Psi_{0}^{c}$ of the lowest state of the
%crystal can be expanded in eigenfunctions $\Psi_{L} $ of the
%operator $\hat{\textbf{L}}^{2}$
%($\hat{\textbf{L}}^{2}\Psi_{L}=\hbar^{2} L(L+1)\Psi_{L}$)
%corresponding to the same energy $E_{0}^{c}$. In view of
%$\hat{\textbf{L}}\Psi_{0}^{c}\neq 0,$ such expansion must contain at
%least one function corresponding to $L\neq 0$. Such state
%$\Psi_{0}^{c}$ is degenerate, since a nonzero $L$ corresponds to
%$2L+1>1$ different states with the energy $E_{0}^{c}$ \cite{land3,vak}.
%If the anisotropic state is characterized by the zero momentum, then
%the degeneration can have a finite or infinite multiplicity,
%generally speaking.

In the 2D case, we have only the operators $\hat{P}_{x},
\hat{P}_{y}, $ and
$\hat{L}_{z}=-i\hbar\sum_{j=1}^{N}\frac{\partial}{\partial
\varphi_{j}}$.  Similarly to the above analysis (replacing in it
$\hat{\textbf{L}}^{2}$ by $\hat{L}_{z}$), we obtain that the states
with nonzero momentum are degenerate. Consider the states with
nonzero $L_{z}$. The replacement $\varphi_{j}\rightarrow
-\varphi_{j}$ for $j=1,\ldots,N$ changes the sign of $\hat{L}_{z}$,
but does not change the 2D Hamiltonian (\ref{1-1}) written in the
cylindrical coordinate system $(\rho,\varphi)$. Therefore, the
states with $L_{z}=\hbar m_{L}$ and $L_{z}=-\hbar m_{L}$ correspond
to the same energy. Thus, only the state, for  which
$\hat{P}_{x}\Psi= \hat{P}_{y}\Psi= \hat{L}_{z}\Psi =0$, is not
degenerate. This is an isotropic translationally invariant state of
the infinite 2D system.

Since any state of the infinite crystal is anisotropic, it is
degenerate. However, it was proved in a monograph \cite{gilbert},
$\S 7$ that the lowest state of the Schr\"{o}dinger problem with
zero BCs is non-degenerate. This was proved for one particle located
in a finite 2D volume. The proof \cite{gilbert} can be easily
generalized to the case of arbitrary $N$ and any dimensionality of
space (see also Appendix 2 in \cite{crys1}). In addition, it was
proved for a one-dimensional (1D) system of $N\geq 2$ point bosons
that each collection of quantum numbers $n_{j}$ (from the Bethe
equations) corresponds to a single solution. This is shown for
periodic \cite{takahashi1999} and zero \cite{mtjpa2017} BCs. Since
GS corresponds to one collection $\{n_{j}\}$
\cite{mtjpa2017,gaudinm}, GS of a one-dimensional system of point
bosons is non-degenerate. These results indicate that GS should be
non-degenerate in the general case: for the finite and infinite
systems, any BCs, $N$, potential, and dimensionality of space.

Because the lowest state ($\Psi_{0}^{c}$) of the 2D or 3D crystal is
degenerate, this is not the genuine GS of the Bose system. By the
theorem of nodes \cite{gilbert},
$\Psi_{0}^{c}(\textbf{r}_{1},\ldots,\textbf{r}_{N})$ must have
nodes. In this case, GS of a liquid is isotropic and translationally
invariant and is described by the nodeless WF
\cite{woo1972,feenberg1974}
\begin{equation}
\Psi_{0} =  C\cdot e^{S_{0}}, \label{1-13}    \end{equation}
\begin{eqnarray}
  && S_{0} =  \sum\limits_{j_{1}<j_{2}}S_{2}(\textbf{r}_{j_{1}}-\textbf{r}_{j_{2}})\nonumber \\
   &&+
\sum\limits_{j_{1}<j_{2}<j_{3}}S_{3}(\textbf{r}_{j_{1}}-\textbf{r}_{j_{2}},\textbf{r}_{j_{2}}-\textbf{r}_{j_{3}})+\ldots
\label{1-14} \\ &&+\sum\limits_{j_{1}<j_{2}<\ldots <
j_{N}}S_{N}(\textbf{r}_{j_{1}}-\textbf{r}_{j_{2}},\textbf{r}_{j_{2}}-\textbf{r}_{j_{3}},\ldots
, \textbf{r}_{j_{N-1}}-\textbf{r}_{j_{N}}).
 \nonumber    \end{eqnarray}
According to the above analysis, such state is non-degenerate.
\textit{Therefore, the genuine nodeless ground state of the infinite
2D or 3D system of spinless bosons is a liquid} for any density and
any repulsive interatomic potential. This is the main result of the
present work. Our conclusion does not related to the Fermi systems,
because for them $\Psi_{0}$ necessarily has nodes \cite{land3}.

%The above-used transition to the infinite system
%is a trick allowing us to make significant conclusions without calculations.
%However, any real system is finite. For such systems, the above analysis is not valid,
%because BCs are not invariant under translations and
%any rotations. We indicate two possibilities: (1)
%the properties of finite and infinite systems differ qualitatively from one another; then the
%true GS of a finite system can be a crystal; (2) the properties of
%finite and infinite systems are similar; then the true GS of a finite
%system corresponds always to a fluid. Case (1) is supported by that the experiment
%and theory deal with finite crystals and, as if,
%aree with (1). However, case (1) contradicts the intuition.
%As for the experiment, we will discuss it at the end of the work. In addition, the exact
%solutions for infinite \cite{ll1963} and finite
%\cite{mt2015,mt1Dcrys1} 1D systems of point bosons indicate that, at the
%transition from the large finite system to the infinite one, no jump in the
%GS energy per atom arises. At the same time, all solutions for 2D and
%3D crystals are approximate. On the whole, we believe that namely case (2) (from two strange possibilities) is realized in the nature.

The real crystals are finite. For a finite system, BCs are not
invariant under the translations. However, BCs can be invariant
under arbitrary rotations, if BCs are set on a circle (for a 2D
system) or on a sphere (for a 3D system).  Then the above analysis
involving the operator $\hat{L}_{z}$ (for a 2D system) or
$\hat{\textbf{L}}^{2}$ (for a 3D system) is valid. The analysis
implies that any anisotropic states of such systems are degenerate.
In addition, it is intuitively clear that the bulk structure of WF
should be independent of the shape of boundaries and should be
invariant at the transition of the large finite system to an
infinite one. In particular, in the last case, no jump of the GS
energy per atom should arise. This is confirmed by the exact
solutions for infinite \cite{ll1963} and finite
\cite{mt2015,mt1Dcrys1} 1D systems of point bosons. Therefore, we
may expect that WF of the lowest state of a finite crystal of any
shape is similar to WF of an infinite crystal and has nodes.

In works \cite{crys1,mt1Dcrys1} it is proved by different methods
that the state of an infinite (or finite ball-shaped) 3D system can
be non-degenerate only in the isotropic case ($L= 0$).

For a  finite 2D or 3D system, it is impossible to find analytically
the exact WF of a crystal, and the numerical methods are not exact
and give the incomplete information. The above-used transition to
the infinite system is a trick allowing us to make important
conclusions without calculations.

Previously, several ans\"{a}tze have been proposed for GS of a
crystal. The localized-Jastrow ansatz is as follows
\cite{saunders1962,brueckner1965,nosanow1966} (see also reviews
\cite{guyer,leggett2006}):
 \begin{equation}
 \Psi^{c}_{0} = C e^{S_{0}}\sum\limits_{P_{c}} e^{\sum_{j=1}^{N}\varphi(\textbf{r}_{j}-\textbf{R}_{j})},
 \label{1-10}    \end{equation}
where $\varphi(\textbf{r})=-\alpha^{2} \textbf{r}^{2}/2$,
$\textbf{r}_{j}$ and $\textbf{R}_{j}$ are the coordinates of atoms
and lattice sites, respectively, $N$ is the number of atoms in the
system, $P_{c}$ means all possible permutations of coordinates
$\textbf{r}_{j}$. Here, $\textbf{R}_j$ are fixed and are the same
for all possible configurations $\{\textbf{r}_{j}\}$.

The wave-Jastrow ansatz reads \cite{woo1976,ceperley1978}:
 \begin{equation}
 \Psi^{c}_{0} = C e^{S_{0}-\sum\limits_{j=1}^{N} \theta(\textbf{r}_{j})},
 \label{1-11}    \end{equation}
where the function $\theta(\textbf{r})$  is periodic with the
periods of a crystal. In ans\"{a}tze (\ref{1-10}) and (\ref{1-11}),
the function $S_{0}$ is usually written in the Bijl--Jastrow
approximation $S_{0} = \sum_{l <
j}S_{2}(\textbf{r}_{l}-\textbf{r}_{j})$ \cite{bijl,jastrow1955}.

Let our system be infinite and periodic. As shown above, the genuine
GS of such a system should be characterized by the zero momentum.
However, it is easy to show that WFs (\ref{1-10}) and (\ref{1-11})
are not eigenfunctions of the operator of total momentum
$\hat{\textbf{P}}= -i\hbar \sum_{j}\frac{\partial}{\partial
\textbf{r}_{j}}$ \cite{crys1}. This difficulty was already noted in
\cite{vitiello1988,reatto1998,reatto2009}, where the translationally
invariant ``shadow''  WF
 \begin{equation}
\Psi_{T}(r)=\int e^{-\Xi(r,s)}ds, \quad
r\equiv\textbf{r}_{1},\ldots,\textbf{r}_{N}
 \label{1-12}    \end{equation}
was proposed instead of (\ref{1-10}). Here,
$s\equiv\textbf{s}_{1},\ldots,\textbf{s}_{N}$ are the ``shadow''
variables, and
$\Xi(r,s)=\sum_{j_{1}<j_{2}}u_{r}(|\textbf{r}_{j_{1}}-\textbf{r}_{j_{2}}|)+
\sum_{k}u_{sr}(|\textbf{r}_{k}-\textbf{s}_{k}|)+\sum_{j_{3}<j_{4}}u_{s}(|\textbf{s}_{j_{3}}-\textbf{s}_{j_{4}}|)$.
Function (\ref{1-12}) is a partial case of the translationally
invariant WF (\ref{1-13}), (\ref{1-14}). The latter sets the general
form of the exact (isotropic or not) translationally invariant WF of
the ground state of the Bose system. Therefore, if the lowest state
of an infinite crystal is described by a nodeless WF, it should take
the form (\ref{1-13}), (\ref{1-14})
\cite{crys1,mcmillan1965,chester1970,reatto1995}.

Thus, only the nodeless ans\"{a}tze were previously considered for
GS of a Bose crystal. However, above we have shown  that the lowest
state of a crystal should be described by WF with nodes.

We remark that though WF (\ref{1-10}) is not  translationally
invariant and is nodeless, it is a reasonable zero approximation for
the lowest  state of the crystal
\cite{crys1,nosanow1966,cazorla2008}.

The nature of the degeneracy of anisotropic solutions is illustrated
by the following examples: the solution
$\Psi=e^{i\textbf{k}\textbf{r}}$ for a free particle, solution
$\Psi_{\textbf{k}}(\textbf{r}_1,\ldots,\textbf{r}_N) =
C_{1}\rho_{-\textbf{k}}\Psi_{0}$ for a Bose liquid with one phonon
\cite{bijl,bz1956,fey1972,yuv2}, and solution
$\Psi_{\textbf{k}_{1}\textbf{k}_{2}}(\textbf{r}_1,\ldots,\textbf{r}_N)
= C_{11}\rho_{-\textbf{k}_{1}}\rho_{-\textbf{k}_{2}}\Psi_{0}$ for a
Bose liquid with two phonons \cite{fey1954,holes2020} (here,
$\Psi_{0} = C\cdot e^{S_{0}}$ and
$\rho_{-\textbf{k}}=N^{-1/2}\sum_{j=1}^{N}e^{i\textbf{k}\textbf{r}_{j}}$).
The solutions for a liquid are written in the zero approximation.
All those solutions are infinitely degenerate, since a rotation of
the vectors $\textbf{k}$, $\textbf{k}_{1}$, and $\textbf{k}_{2}$
transfers them into another solutions with the same energies
(vectors $\textbf{k}_{1}$ and $\textbf{k}_{2}$ must be rotated by
the same angle). The direct solving of the Schr\"{o}dinger equation
shows this. On the other hand, such solutions are characterized by
nonzero momentum and, therefore, should be degenerate according to
the analysis at the beginning of the present work. All such one- and
two-phonon solutions contribute to the statistical sum and the heat
capacity. That is, they are physically different states. We now
write the following $6J$-phonon solution in the zero approximation:
 \begin{eqnarray}
    \Psi_{0}^{c}(\textbf{r}_1,\ldots,\textbf{r}_N) =
 C(\rho_{\textbf{k}_{x}}\rho_{-\textbf{k}_{x}}\rho_{\textbf{k}_{y}}\rho_{-\textbf{k}_{y}}\rho_{\textbf{k}_{z}}\rho_{-\textbf{k}_{z}})^{J}e^{S_{0}},
       \label{c-p}\end{eqnarray}
where $\textbf{k}_{x}=\textbf{i}_{x}(\textbf{k}_{r}\textbf{i}_{x})$,
$\textbf{k}_{y}=\textbf{i}_{y}(\textbf{k}_{r}\textbf{i}_{y})$,
$\textbf{k}_{z}=\textbf{i}_{z}(\textbf{k}_{r}\textbf{i}_{z})$.
$\textbf{k}_{r}$ is a vector of the reciprocal lattice (with nonzero
smallest components) of a 3D crystal with a rectangular lattice.
Note that the nodal structure of WF for a state with a given
collection of phonons is independent of the coupling  constant (for
details see \cite{mt1Dcrys1}). Therefore, the structure of a binary
correlation function $g_{2}(\textbf{r}_{1},\textbf{r}_{2})$ at a
strong coupling can be seen by calculating
$g_{2}(\textbf{r}_{1},\textbf{r}_{2})$ at zero coupling (free
particles) for the same ensemble of phonons. It is known that when
the interatomic interaction is switched-off, the phonons are
transformed into free bosons with the same momenta \cite{holes2020}.
Respectively, WF (\ref{c-p}) is transformed into WF of free bosons
$|\Psi_{0}^{c}\rangle =
|N_{\textbf{k}_{x}},N_{-\textbf{k}_{x}},N_{\textbf{k}_{y}},N_{-\textbf{k}_{y}},N_{\textbf{k}_{z}},N_{-\textbf{k}_{z}},N_{\textbf{k}=0}\rangle$,
where $N_{\textbf{k}}$ is the number of bosons with momentum
$\hbar\textbf{k}$. In our case,
$N_{\textbf{k}_{x}}=N_{-\textbf{k}_{x}}=N_{\textbf{k}_{y}}=N_{-\textbf{k}_{y}}=N_{\textbf{k}_{z}}=N_{-\textbf{k}_{z}}=J$,
$N_{\textbf{k}=0}=N-6J$. Setting
 \begin{eqnarray}
&&\hat{\psi}(\textbf{r},t) =V^{-1/2}\left
(\hat{a}_{\textbf{k}_{x}}e^{ik_{x}x}+\hat{a}_{-\textbf{k}_{x}}e^{-ik_{x}x}+
\hat{a}_{\textbf{k}_{y}}e^{ik_{y}y}\right.\nonumber \\ &&+ \left.
\hat{a}_{-\textbf{k}_{y}}e^{-ik_{y}y}+
\hat{a}_{\textbf{k}_{z}}e^{ik_{z}z}+\hat{a}_{-\textbf{k}_{z}}e^{-ik_{z}z}+\hat{a}_{0}\right
),
       \label{Psi-n}\end{eqnarray}
after some calculations we find
 \begin{eqnarray}
  && g_{2}(\textbf{r}_{1},\textbf{r}_{2}) \nonumber \\ && \equiv  C_{g}\langle \Psi_{0}^{c}|\hat{\psi}^{+}(\textbf{r}_{1},t)\hat{\psi}^{+}(\textbf{r}_{2},t)
\hat{\psi}(\textbf{r}_{1},t)\hat{\psi}(\textbf{r}_{2},t)|\Psi_{0}^{c}\rangle
 \label{g2-1} \\ && =
 C_{g}V^{-2}\left [N^{2}-N-6J^{2}+4JF_{3}\cdot(JF_{3}+N-6J)\right ],
       \nonumber \end{eqnarray}
where
 \begin{eqnarray}
 F_{3}&=&\cos{[k_{x}(x_{1}-x_{2})]}+\cos{[k_{y}(y_{1}-y_{2})]}
\nonumber \\ &+&\cos{[k_{z}(z_{1}-z_{2})]},
       \label{Psi-n}\end{eqnarray}
$C_{g}=V^{2}/(N^{2}-N)$. It is a translationally invariant
crystal-like solution with a rectangular 3D lattice. It is natural
to expect that the ideal crystal corresponds to
$g_{2}(\textbf{r}_{1},\textbf{r}_{2})$  with the largest amplitude
of oscillations, which leads to $J=N/6$ (this can be easily seen
with regard for the condition $6J\leq N$ following from that the
total number of phonons cannot exceed the number of atoms
\cite{holes2020}). On this basis, we suppose that WF (\ref{c-p})
with $J=N/6$ is the zero approximation for WF of the ground state of
a crystal with the rectangular lattice for a realistic (not weak)
interatomic interaction. We note that $|\Psi_{0}^{c}\rangle =
|N_{\textbf{k}_{x}},N_{-\textbf{k}_{x}},N_{\textbf{k}_{y}},N_{-\textbf{k}_{y}},N_{\textbf{k}_{z}},N_{-\textbf{k}_{z}},N_{\textbf{k}=0}\rangle$
is the exact WF of a system of free bosons. In the coordinate
representation, it is equal to the sum of the main summand
$C(\rho_{\textbf{k}_{x}}\rho_{-\textbf{k}_{x}}\rho_{\textbf{k}_{y}}\rho_{-\textbf{k}_{y}}\rho_{\textbf{k}_{z}}\rho_{-\textbf{k}_{z}})^{J}$
and small corrections with fewer variables $\rho_{\textbf{k}}$. In
the formalism of second quantization, it is not necessary to know
them.

On the other hand, WF (\ref{c-p}) describes  a liquid containing $J$
identical phonons with momenta $\textbf{k}_{x}, -\textbf{k}_{x},
\textbf{k}_{y}, -\textbf{k}_{y}, \textbf{k}_{z}, -\textbf{k}_{z}$.
This is the state with six condensates of phonons. The crystal-like
$g_{2}(\textbf{r}_{1},\textbf{r}_{2})$ indicates that such a
composite condensate of quasiparticles creates a crystal lattice. It
is obvious that if we turn the vectors $\textbf{k}_{x}$,
$\textbf{k}_{y}$, and $\textbf{k}_{z}$ in $\Psi_{0}^{c}$ (\ref{c-p})
through the same arbitrary angle $\varphi \textbf{i}_{\varphi}$,  at
$N, V=\infty$ we obtain another solution of our boundary-value
problem, corresponding to the same energy. In other words, such
state is infinitely degenerate. Note that solution (\ref{c-p})
always exists, since it is merely the solution for of a liquid with
$6J$ interacting phonons.

%The above-presented one-, two- and multiphonon solutions belong to the
%complete collection of eigenfunctions of the operators $\hat{H}$ and $\hat{\textbf{P}}$.

It is interesting to compare these properties with properties of the
solution for a hydrogen atom, $\psi(\textbf{R},\textbf{r})=const
\cdot e^{i\textbf{P}\textbf{R}/\hbar}\psi_{nlm}(\textbf{r})$, where
$\textbf{R}$ is the coordinate of the center of masses, and
$\textbf{r}$ is the coordinate of the electron relative to the
nucleus \cite{land3,vak}. If we turn the vector $\textbf{P}$ in
$e^{i\textbf{P}\textbf{R}/\hbar}$ through any  angle, we get another
solution with the same energy. But if the function
$\psi_{nlm}(\textbf{r})$ will be ``turned'' through any  angle, we
do not obtain again a solution (except for the case of $l=0$). Why?
The function $e^{i\textbf{P}\textbf{R}/\hbar}$ corresponds to the
Hamiltonian $\hat{H}_{1}=\hat{\textbf{P}}^{2}/2M$ that commutes with
the operator of momentum of the center of masses $ \hat{\textbf{P}}$
and the operator of angular momentum $[\textbf{R}\times
\hat{\textbf{P}}]$. The function $\psi_{nlm}(\textbf{r})$
corresponds to $\hat{H}_{2}=\hat{\textbf{p}}^{2}/2m+U(r)$ that
commutes with  the operator $\hat{\textbf{l}}=[\textbf{r}\times
\hat{\textbf{p}}]$ and \textit{does not} commutes with
$\hat{\textbf{p}}$. Therefore, the infinite-fold degeneracy that
reveals itself in the many-boson problem and is related to the
noncommutativity of the operators $\hat{\textbf{L}}$ and
$\hat{\textbf{P}}$ is not inherent in WF of the electron
$\psi_{nlm}(\textbf{r})$.  In this case, the $(2l+1)$-fold
degeneracy of the state $\psi_{nlm}(\textbf{r})$  is caused by that
the operators $\hat{l}_{x}, \hat{l}_{y}$ and $\hat{l}_{z}$ do not
commute with each other and commute with $\hat{H}_{2}$
\cite{land3,vak}.

The above analysis involves rotations and is not applicable to the
1D case. According to the exact crystalline  solution for a 1D
system of point bosons with zero BCs \cite{mt1Dcrys1},  WF of the
lowest state of the crystal contains nodes. However, for the
long-range repulsive potential, the genuine GS of a finite system
can be a crystal
\cite{lozovik2005,reimann2010,zollner2011,zollner2011b,chatterjee2013,chatterjee2019,lode2020}.

The authors of many Monte Carlo simulations for 2D and 3D Bose
systems with $N\sim 100$  reported on crystal-like solutions with a
presumably nodeless WF (see reviews
\cite{cazorla2017,reatto1995,ceperley1992,kalos2008} and Appendix 3
in \cite{crys1}). In this case in the majority of works, the
functions $g_{2}(\textbf{r}_{1}-\textbf{r}_{2})$ and $S(\textbf{k})$
that could show the anisotropy corresponding to a crystal were not
calculated. We assume that those crystalline solutions correspond to
WFs with nodes, since the above analysis shows that any crystal-like
WF of a macroscopic system should have nodes. The future studies
will clarify why the Monte Carlo method did not catch this property
for many years.

Finally, consider the experiment. Our analysis shows that the
genuine nodeless GS of a Bose system of any density $\rho$ must
correspond to a liquid. Therefore, the following inequality should
hold at any $\rho$:
\begin{eqnarray}
   E_{0}^{c}(\rho, N) >  E_{0}^{l}(\rho, N).
       \label{e00}\end{eqnarray}
Here, $E_{0}^{c}$ and $E_{0}^{l}$ are the energies of the lowest
states of a Bose crystal and a Bose liquid, respectively. Inequality
(\ref{e00}) evidences that there should exist a region of states of
a liquid, the energy of which is less than the GS energy of a
crystal of the same density.
% Inequality (\ref{e00}) evidences that
%a region of liquid states, whose energies are less than the GS
%energy of a crystal of the same density, should exist.
It is natural to call a liquid in such states the ``under-crystal
liquid''. It is clear that, at very low temperatures, it should be
superfluid. In experiments, one such liquid has already been
obtained: it is He II. Due to large zero oscillations of atoms, the
solid $^4$He is unstable at $P< 25$ atm. Other inert elements are
characterized by smaller zero oscillations
\cite{cazorla2017,debour1948} and have a stable crystalline phase at
low $P, T$. One can transform them into the state of under-crystal
liquid, apparently, in several ways \cite{crys1}: (i) by overcooling
the liquid at $P>P_{3}$; (ii) by compressing a gas whose state on
the $(P,T)$ diagram lies lower than the sublimation curve; (iii) by
applying a negative pressure to the crystal at $T<T_{3}$ ($P_{3}$
and $T_{3}$ are the pressure and temperature at the triple point).

We suppose that the under-crystal liquid has not been obtained
earlier in experiments with H$_{2}$, Ne, Ar, Kr, and Xe because of
peculiarities of the formation of nuclei. Perhaps, the Nature
deceives us in such a way. The main point is that methods (i) and
(ii) under the ordinary conditions must lead to the formation of
crystal nuclei and to the crystallization.  This is because the
medium contains impurities, and the walls of the vessel contain
defects.  The walls and impurity particles have usually a
crystalline structure, which decreases the work of creation of
crystal nuclei \cite{volmer,kashchiev} and stimulates the
crystallization. Therefore, it is important for methods (i) and (ii)
to create the special conditions that will prevent the formation of
crystal nuclei. Such conditions are discussed in \cite{crys1} in
detail. If we would manage to prevent the crystallization, our
probable prize will be getting liquid superfluid H$_{2}$, Ne, Ar,
Kr, and Xe in macroscopic amounts.

Some of these  liquids can be metastable. The reason is as follows.
The crystal and liquid are in the equilibrium at the same pressure
\cite{gibbs}. Inequality (\ref{e00}) admits $E_{0}^{c}(P)
> E_{0}^{l}(P)$ and $E_{0}^{c}(P) < E_{0}^{l}(P)$. In the first case,
the liquid is stable relative to the crystallization. In the second
case, it is metastable and can be crystallized. For $^4$He, namely
the inequality $E_{0}^{c}(P) > E_{0}^{l}(P)$ holds at the melting
point (see \cite{crys1} and references therein). It is natural to
expect that  $E_{0}^{c}(P)
> E_{0}^{l}(P)$ also for several other inert elements. Then the corresponding
under-crystal liquids must be stable relative to the
crystallization.

Thus, the general quantum mechanical analysis allows us to make some
conclusions about the nature of 2D and 3D Bose crystals.  We may
expect that the future research will lead to interesting results.

The present work is partially supported by the National Academy of
Sciences of Ukraine (project No.~0121U109612).

%Dear Editors,

%the main result of the work is explained in the Abstract. In my
%opinion, the manuscript deserves to be published in Phys. Rev. Lett.
%Thank you.

%Yours sincerely, Maksim Tomchenko

       \end{document}